\documentclass[aps,prl,twocolumn,floatfix,superscriptaddress,amsmath,amssymb]{revtex4-2}
\usepackage{graphicx}
\usepackage{bm}
\newcommand{\rhead}[1]{\par\textit{#1}---\ignorespaces}
\newcommand{\vk}{\mathbf{k}}

\begin{document}

\title{Electron versus positron impact ionization of inner shells of heavy atoms: distortion, exchange and relativistic effects}
\author{Colm T. Whelan}
\affiliation{Department of Physics, Old Dominion University, Norfolk, Virginia 23529, USA}

\begin{abstract}
Triple differential cross sections are presented for electron and positron impact ionization of the K shells of Cu, Ag, Au and U and the L shells of Ag. Relativistic effects are important throughout. For positrons the recoil maximum is suppressed and the recoil-region spin asymmetry changes sign; for the $2s$ shell the binary peak changes. In symmetric geometry, where theory reproduces the measured absolute electron cross section, the positron peak lies on the other side of zero recoil. Measurements would test relativistic ionization theory in a regime that electron experiments cannot reach.
\end{abstract}

\maketitle

\begin{figure}
\includegraphics[scale=0.25]{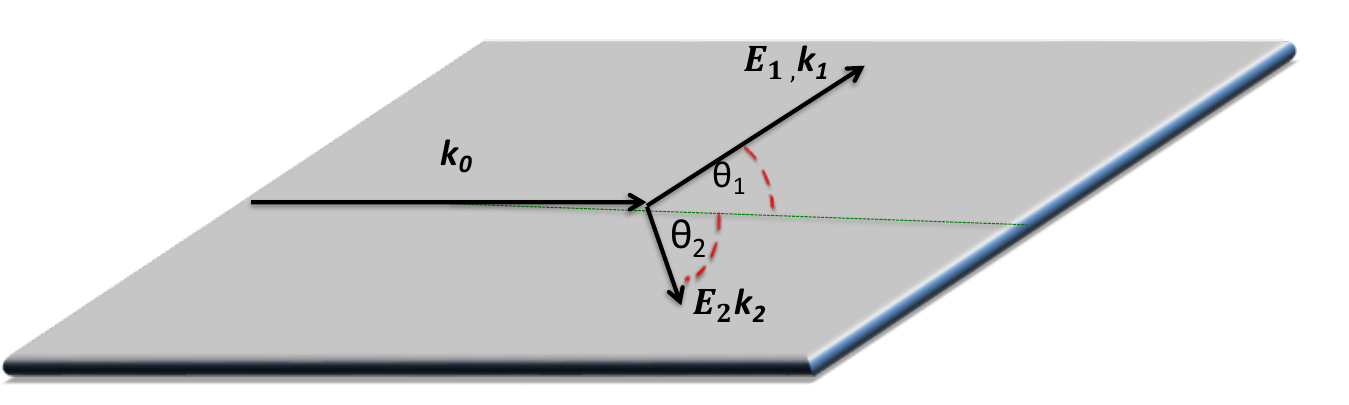}
\vspace*{-1em}
\caption{In a coincidence experiment an incoming projectile ionizes an atom and the scattered projectile and ejected electron are detected with their energies and angles resolved. In this paper only cases where all  three vectors $\vk_0,\vk_1$, and $\vk_2$,  lie in the same plane. Both coplanar asymmetric geometry: $\theta_1$ fixed, $ E_1\gg E_2$, with the TDCS given as a function of $\theta_2$ and coplanar symmetric, $E_1=E_2$, $\theta_1=\theta_2=\theta $ are considered.\label{coplanar}}
\end{figure}

\rhead{Introduction}
A coincidence experiment, in which the energies and directions of both outgoing particles are
fixed, is the most stringent test an ionization theory can face. It probes the collision amplitude
itself rather than an average over unobserved variables, and by choosing the kinematics the
experimenter can steer the collision into regions where a single, subtle mechanism dominates, one
whose signature would be suppressed in a less differential measurement (Fig.~ \ref{coplanar}). 
Nakel and his collaborators have performed an exhaustive study of electron impact ionization of the K and L shells of heavy atom targets\cite{nakel99}.
The impinging electrons were traveling at a significant proportion of the speed of light: 300\,keV ($v_0 \approx0.78c$) and 500\,keV ($v_0\approx 0.86c$).
 The  Nakel measurements  were made on thin solid foils. Because the areal
density of a foil can be measured directly, these experiments determined the absolute size of the
cross section accurately, unlike low-energy gas-target $(e,2e)$ experiments, which are usually put on
an absolute scale only by normalization to theory or to other data \cite{rasch97}. In the relativistic case
theory was  therefore  challenged  on magnitude as well as shape. 
In response to these absolute measurements a full relativistic theory was developed, together with an associated code (rDWBA).   
It  has been  shown that the triple differential cross section
(TDCS) is shaped by the distortion of all three continuum electrons in the field of the nucleus, so
that plane-wave theories fail even at several hundred keV \cite{walters92,keller94a}; that a recoil
maximum, in which the nucleus absorbs the momentum transfer, grows with the nuclear charge until for
gold and uranium it rivals the binary peak \cite{keller94a,keller96}; that in symmetric geometry the
binary peak is displaced from zero recoil momentum, a purely relativistic effect that the magnetic and
retarded parts of the electron--electron interaction leave on the cross section and that the
experiments were not designed to find \cite{ast96,nakel99}; and that projectiles polarized
perpendicular to the scattering plane produce a spin up--down asymmetry of up to 40\%
\cite{prinz95,keller96}. In the non-relativistic limit spin and orbital angular momentum are
separately conserved and this asymmetry vanishes; it is a direct signature of the spin--orbit
interaction of the continuum electrons with the nucleus, that is, of Mott scattering inside an
ionizing collision \cite{nakel99}.
Positrons have so far been excluded from this programme because no sufficiently intense beam
existed. That is changing. Moderated positron beams of high intensity are now available
\cite{hugenschmidt16}, and $(e^+,e^+e^-)$ coincidence experiments have been carried out at low
energies \cite{kover98,arcidiacono05,dubois21}. At Jefferson Laboratory the PEPPo experiment has
produced positrons with polarization up to 82\% from a polarized electron beam of only a few MeV
\cite{abbott16}, and a programme of polarized and unpolarized positron beams of high duty cycle is
being prepared \cite{accardi21}. Coincidence experiments with fast positrons on inner shells are
therefore coming within reach. Given the wealth of physics uncovered with electrons, it is
important that a reliable theoretical framework be in place before such experiments are designed,
to show where to look and what can be learned. The natural first step is to ask what is gained by comparing positron and electron impact in the
same kinematical arrangement. Two
things change when the electron is replaced by a positron. First, distortion by the atomic field
depends on the sign of the charge: the electron is attracted towards the nucleus and the positron
repelled, and the spin--orbit force changes sign. Second, the positron is distinguishable from the
atomic electrons, so exchange is switched off, both between the two outgoing particles and in the
elastic scattering of the positron by the atom. Because the ejected particle is an electron in
both cases, every difference between electron and positron impact can be traced to these two
effects. In this paper it is shown that the switch from electron to positron makes a significant change to the TDCS:
reversing the charge of the projectile  can all but remove the recoil maximum,
can more than double the binary peak and reverse the sign of the spin asymmetry.

\rhead{Theory}
In this letter the relativistic distorted-wave Born approximation as described in
Refs.~\cite{keller94a,nakel99,keller94} is used.  The rDWBA  code was revised and improved not only to allow for positron scattering but also to give improved convergence.
For unpolarized particles the TDCS is \cite{nakel99}
\begin{widetext}
\begin{equation}
\frac{d^3\sigma}{d\Omega_1 d\Omega_2 dE_2}
=\frac{(2\pi)^2}{c^6}\frac{k_1k_2}{k_0}E_0E_1E_2
\frac{N_\kappa}{2N_m}\sum_{\epsilon_0\epsilon_1\epsilon_2\epsilon_b}\left|S^{\rm dir}-S^{\rm ex}\right|^2 ,
\label{eq:tdcs}
\end{equation}
\end{widetext}
where 0, 1, 2 and b label the incident, the two outgoing and the bound particle, $E_i$ and $k_i$ are
total energies and momenta, $\epsilon_i$ spin projections, and $N_\kappa/N_m$ is the occupation of the
bound subshell $\kappa$ divided by its degeneracy. In the first-order M{\o}ller approximation the
$S$-matrix element is the difference of a direct and an exchange term, the direct term being
\begin{widetext}
\begin{equation}
S^{\rm dir}
=i\!\int\! d^4x\, d^4y\,\bar\psi_{k_1\epsilon_1}(x)\gamma^\mu\psi_{k_0\epsilon_0}(x)\,
D^0_{\mu\nu}(x-y)\,\bar\psi_{k_2\epsilon_2}(y)\gamma^\nu\psi_{\kappa\epsilon_b}(y) ,
\label{eq:Sdir}
\end{equation}
\end{widetext}
with the free photon propagator
$D^0_{\mu\nu}(x-y)=-(4\pi i/c)\,g_{\mu\nu}\int d^4q\,e^{iq(x-y)}/[(2\pi)^4(q^2+i\varepsilon)]$, which
after the time integrations gives the retarded interaction
$(1-\bm{\alpha}_1\!\cdot\!\bm{\alpha}_2)\,e^{i\omega r_{12}/c}/r_{12}$, $\omega=E_0-E_1$; the
exchange term $S^{\rm ex}$ follows from $S^{\rm dir}$ by interchanging $(k_1\epsilon_1)$ and $(k_2\epsilon_2)$. All
continuum orbitals $\psi$ are exact solutions of the Dirac equation in a central potential, were
computed with the program of Salvat and Mayol \cite{salvat91}; final-state repulsion between the two
outgoing electrons is neglected \cite{nakel99}. Exchange thus enters in two places: between the two
final electrons, through $S^{\rm ex}$, and in the elastic scattering of the continuum electrons by the atom,
which is described by the local exchange--correlation part of the potential. As in
Ref.~\cite{keller94a} the potential is the self-consistent relativistic Kohn--Sham local-density
potential of the neutral atom (nuclear, Hartree and local exchange--correlation terms); the original
numerical potentials have been replaced  with \textsc{dftatom} \cite{certik13}. The same
potential is used for the incident, scattered and ejected electrons. The K-shell orbital is, as in
Ref.~\cite{keller94a}, the Dirac hydrogenic $1s_{1/2}$ function for the full nuclear charge; the L-shell
orbitals ($2s_{1/2}$, $2p_{1/2}$, $2p_{3/2}$) are the Dirac orbitals of the self-consistent potential.
Experimental binding energies are used in the kinematics. The  results of the revised  rDWBA program  agrees with the earlier code to within a few percent over the full range of the earlier calculations.

For positron impact $S^{\rm ex}=0$, and the incident and scattered waves are calculated in the static
(nuclear plus Hartree) potential with its sign reversed; the local exchange--correlation term, which
describes electron--electron exchange, is omitted. The ejected electron
is treated exactly as for electron impact. Positronium formation and real annihilation are
negligible at these energies. The virtual-annihilation amplitude is not included; for free
$e^+e^-$ scattering at 500\,keV it lowers the cross section by 12\% when the slower electron has
100\,keV and by 28\% at equal energy sharing, so it affects absolute positron cross sections at the
level of the uncertainties quoted here, and it will be added in future work. The multipole expansion
of the electron--electron interaction is taken to order 7 in asymmetric kinematics, where higher
orders change the cross section by less than 2\%, and to order 20 in symmetric kinematics, where the
ejected electron is fast. The spin asymmetry is
$A=(\sigma_\uparrow-\sigma_\downarrow)/(\sigma_\uparrow+\sigma_\downarrow)$, where the arrows refer
to projectile spin perpendicular to the scattering plane; it is obtained from the amplitudes for the
two helicity states of the projectile.

\rhead{Results}
\begin{figure*}
\includegraphics[width=\textwidth]{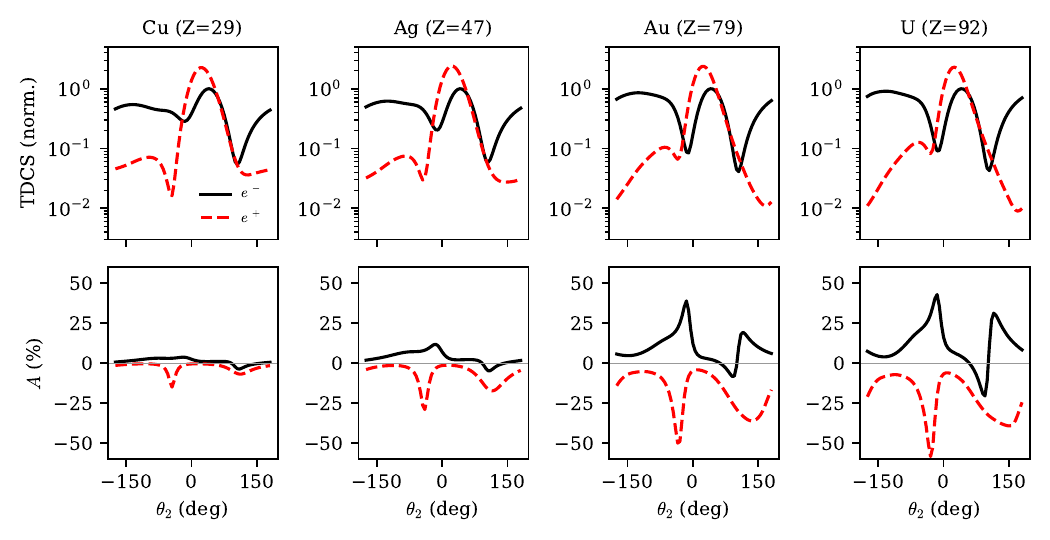}
\caption{K-shell ionization of Cu, Ag, Au and U by electrons (full curves) and positrons (dashed
curves) in the scaled kinematics of Refs.~\cite{keller96,nakel99}: $E_0=3.72\,I_K$, $E_1/E_2=3$,
$\theta_1=-10^\circ$, coplanar geometry ($E_0=300$\,keV for Au). Top: TDCS, normalized to the
electron binary maximum. Bottom: spin up--down asymmetry for projectiles polarized perpendicular to
the scattering plane.}
\label{fig:z}
\end{figure*}

Numerical values for all the curves in Figs.~\ref{fig:z}--\ref{fig:sym} are given in the
Supplemental Material \cite{whelan26}. Figure~\ref{fig:z} compares electron and positron impact on four targets in the scaled kinematics of
Refs.~\cite{keller96,nakel99}. For electrons the recoil maximum grows with $Z$ from 54\% to 91\% of the
binary maximum. For positrons it falls to 5--9\%, a suppression by a factor of 6 (Cu) to 20 (U), and
it moves from $\theta_2\approx-130^\circ$ to $-25^\circ$; the binary peak is 2.3--2.4 times higher for
positrons at all four targets. The behaviour is what one expects if the recoil maximum is produced by
the projectile: the attracted electron can transfer momentum to the nucleus in a close collision,
the repelled positron cannot. At the experimental kinematics of Ref.~\cite{keller94a} for gold
($E_0=500$\,keV, $E_2=100$\,keV, $\theta_1=-15^\circ$) the recoil region is reduced by a factor
5--20 and the binary peak is 2.5 times larger for positrons.

The lower panels show the spin asymmetry. For electrons it is positive throughout the recoil region,
with a maximum near $\theta_2=-15^\circ$ that grows from 4\% (Cu) and 12\% (Ag) to 39\% (Au) and 43\%
(U) \cite{keller96}. For positrons it is negative there, with minima of $-15\%$, $-29\%$, $-50\%$ and
$-58\%$ near $\theta_2=-35^\circ$; it is also negative, by 20--40\% for the heavy targets, at large
angles on the binary side. The same reversal occurs at the experimental kinematics: for silver at
the Prinz kinematics $A$ in the recoil region changes from $+9\%$ to between $-3\%$ and $-14\%$, and
for gold at 500\,keV from $+14\%$ to $-20\%$ at $\theta_2=-40^\circ$. Because the ejected electron is the same in both cases,
the reversal shows that the recoil-region asymmetry is produced by the spin--orbit interaction of the
\emph{projectile} with the nucleus: the sign of the spin--orbit force follows the sign of the charge.

\begin{figure*}
\includegraphics[width=\textwidth]{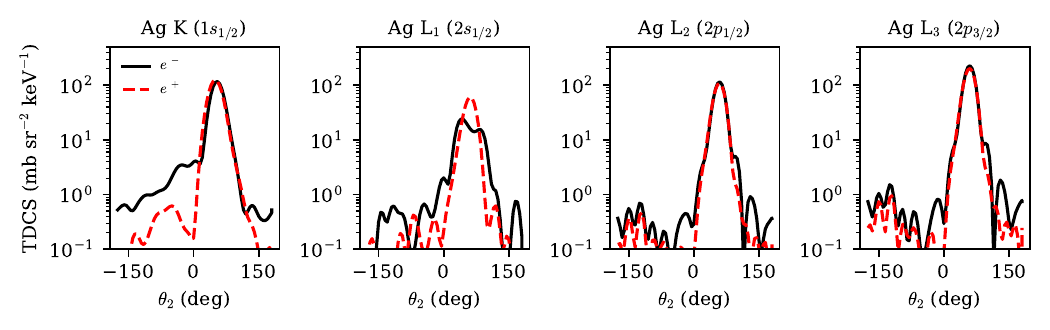}
\caption{Ionization of the K, L$_1$, L$_2$ and L$_3$ shells of silver by electrons (full curves) and
positrons (dashed curves): $E_0=500$\,keV, $E_2=100$\,keV, $\theta_1=-15^\circ$, coplanar geometry
\cite{keller94a}.}
\label{fig:L}
\end{figure*}

Figure~\ref{fig:L} shows the K and L shells of silver at the kinematics of Ref.~\cite{keller94a}. At
this higher energy the binary peak for the K shell is nearly the same for both projectiles, but the
recoil region is still reduced by a factor of about 6 for positrons. The $L_1$ ($2s$) shell behaves
differently: its momentum distribution has a node close to the momentum transfer at the binary peak,
so the cross section there depends strongly on the distortion of the projectile wave. For positrons
the binary maximum is 2.5 times larger than for electrons and moves from $\theta_2=40^\circ$ to
$60^\circ$. For the $2p$ shells, where the recoil region is already weak for electrons, the two
projectiles give similar results.

\begin{figure}
\includegraphics[width=\columnwidth]{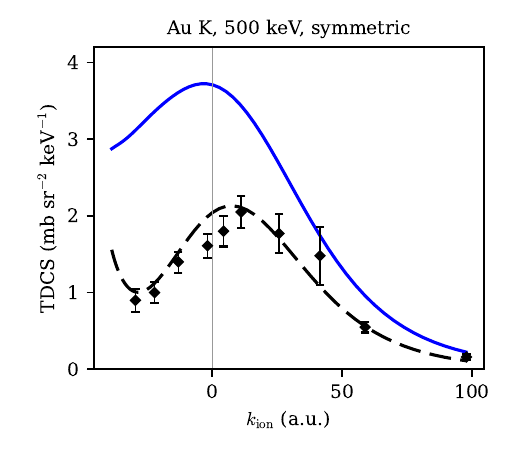}
\caption{K-shell ionization of gold in coplanar symmetric geometry, $E_0=500$\,keV,
$E_1=E_2=209.65$\,keV, as a function of the recoil momentum $k_{\rm ion}$. Dashed curve: rDWBA for
electrons; full curve: rDWBA for positrons. Points: absolute measurement of Bonfert
\emph{et al.}\ \cite{bonfert91} for electrons.}
\label{fig:sym}
\end{figure}

The third case is the coplanar symmetric geometry of Bonfert \emph{et al.}\ \cite{bonfert91}, where the
binary peak for gold is displaced from zero recoil momentum towards positive $k_{\rm ion}$, a purely
relativistic effect \cite{ast96,nakel99}. Our electron calculation reproduces the measurement on an
absolute scale (Fig.~\ref{fig:sym}): the binary maximum is 2.13\,mb\,sr$^{-2}$\,keV$^{-1}$
(measured 2.05) and lies at $k_{\rm ion}=+7.5$ a.u. (measured about $+10$ a.u.). For positrons the
peak moves to the opposite side of zero recoil, $k_{\rm ion}=-3.2$ a.u., and is about 1.75 times higher;
this ratio is an upper estimate, since the virtual-annihilation amplitude, which is largest at equal
energy sharing, is not included.

The displacement of the binary peak is a signature of the relativistic electron--electron
interaction. Besides the Coulomb term, the interaction $(1-\bm{\alpha}_1\!\cdot\!\bm{\alpha}_2)\,
e^{i\omega r_{12}/c}/r_{12}$ that follows from the photon propagator in Eq.~(\ref{eq:Sdir}) contains the
magnetic (current--current) interaction $\bm{\alpha}_1\!\cdot\!\bm{\alpha}_2$ and retardation,
neither of which has a nonrelativistic counterpart. Table~\ref{tab:peak} shows the position of the
peak as the ingredients are switched off. With the instantaneous Coulomb interaction the electron
peak lies just below zero recoil, as found in Ref.~\cite{ast96}, and exchange hardly matters.
Restoring the magnetic and retarded terms moves the peak towards positive recoil for \emph{both}
projectiles, by 4.4\,a.u.\ for the electron direct amplitude and by 8.3\,a.u.\ for the positron. For
electrons the displacement is then more than doubled through the interference of direct and exchange
amplitudes, from $+1.5$ to $+7.6$\,a.u., while distortion of the electron contributes little (the peak
is at $+0.9$\,a.u.\ with a plane-wave projectile). The positron has no exchange amplitude, and its
repulsive distortion pulls the peak back, from $+0.9$\,a.u.\ with a plane-wave projectile to
$-3.4$\,a.u. In the electron case the magnetic interaction acts largely through exchange; with
positrons it competes only with distortion, so that symmetric $(e^+,e^+e^-)$ measurements would
probe the relativistic interaction more directly.

\begin{table}
\caption{Position of the binary maximum, $k_{\rm ion}$ (a.u.), for the K shell of gold in the
geometry of Fig.~\ref{fig:sym}, with multipoles up to order 12 (at order 20 the full results are
$+7.5$ and $-3.2$\,a.u.). DW: distorted projectile; PW: plane-wave projectile; the ejected electron
is always distorted.}
\label{tab:peak}
\begin{ruledtabular}
\begin{tabular}{llcc}
Interaction & Projectile & $e^-$ & $e^+$\\
\hline
Coulomb, no exchange & DW & $-2.9$ & $-11.7$\\
Coulomb, with exchange & DW & $-2.7$ & \\
Full, no exchange & PW & $+0.9$ & $+0.9$\\
Full, no exchange & DW & $+1.5$ & $-3.4$\\
Full, with exchange & DW & $+7.6$ & \\
\end{tabular}
\end{ruledtabular}
\end{table}

Exchange is at its largest in this geometry; in strongly asymmetric kinematics it is much less important,
and for silver at the kinematics of Ref.~\cite{keller94a} it changes the binary peak by less than
1\% and the cross section elsewhere by at most about 10\%.

\rhead{Conclusion}
Comparing electron and positron impact separates the role of the projectile's interaction with the
nucleus in relativistic inner-shell ionization. In the rDWBA the recoil maximum is largely switched
off and the recoil-region spin asymmetry changes sign when the projectile charge is reversed; for
the $2s$ shell the binary peak itself is changed. In symmetric geometry, where the rDWBA reproduces the
measured absolute cross section for electrons, the binary peak for positrons lies on the opposite
side of zero recoil: the magnetic and retarded interaction displaces the peak of both projectiles
towards positive recoil, but for electrons the displacement is amplified by exchange, while for
positrons it is outweighed by the repulsive distortion. A measurement of either effect
would test the relativistic theory of ionization in a regime that electron experiments cannot reach. Given current experimental capabilities, it should be a relatively
straightforward matter to test the predictions made in this Letter.

\rhead{Acknowledgments}The computer code used in this paper is a revised version of the earlier rDWBA electron code
developed by the author and his collaborators \cite{keller94a,nakel99}. It was the author's privilege
to work with the late Professor R. M. Dreizler, the late Professor H. R. J. Walters and Drs Ancarani, Ast,
Allan, Kampp and Keller in developing the original code.

\rhead{Data availability}The data that support the findings of this Letter are available from
the author upon reasonable request.

\bibliography{epm_letter}

\end{document}